\documentclass[10pt,letterpaper,twocolumn]{article}
\usepackage{graphicx}
\usepackage{amsmath}
\usepackage[top=1in, bottom=1.25in, left=0.75in, right=0.75in]{geometry}

\title{Generalized projection retrieval of dispersion scans for ultrashort pulse characterization}
\author{
Miguel~Miranda,\textsuperscript{1,*}
Jo{\~a}o~Penedones,\textsuperscript{2,3}
Chen~Guo,\textsuperscript{1}
Anne~Harth,\textsuperscript{1}\\
Ma{\"i}t{\'e}~Louisy,\textsuperscript{1}
Lana~Neori\v{c}i\'{c},\textsuperscript{1,4}
Anne~L'Huillier,\textsuperscript{1} and
Cord~L.~Arnold\textsuperscript{1}\\
\\
\textsuperscript{1}Department of Physics, Lund University, P.O. Box 118, SE-221 00 Lund, Sweden\\
\textsuperscript{2}Fields and Strings Laboratory, Institute of Physics, EPFL, CH-1015 Lausanne,
	Switzerland\\
\textsuperscript{3}Centro de F\'{i}sica do Porto, Universidade do Porto, Rua do Campo Alegre 687, 4169-007 Porto, Portugal\\
\textsuperscript{4}ELI-ALPS, Dugonics t\'{e}r 13, 6720 Szeged, Hungary\\
\\
\textsuperscript{*}Corresponding author: miguel.miranda@fysik.lth.se
}

\def\F#1{\mathcal{F}\left\{ {#1} \right\}}
\def\invF#1{\mathcal{F}^{-1}\left\{ {#1} \right\}}
\begin{document}

\maketitle

\begin{abstract}
	
	We present a new retrieval algorithm based on generalized projections for ultrashort pulse characterization using dispersion scan (d-scan).  The proposed algorithm is much faster and leads to a drastic reduction of retrieval times but, compared to the standard algorithm, it performs less robustly in the retrieval of noisy d-scan traces. The algorithm is tested on several simulated cases and in two different experimental cases in the few-cycle regime.

\end{abstract}

\section{Introduction}

The last decades have seen the development and improvement of numerous ultrashort laser pulse characterization techniques. Still, temporal characterization of ultrashort laser pulses can be challenging, especially in the few-cycle regime.  Options range from classic techniques, such as FROG~\cite{TrebinoJOSAA1993} and SPIDER~\cite{IaconisOL1998} to more complex techniques based on photo ionization~\cite{ItataniPRL2002,MairessePRA2005,WyattO2016}, which allow for direct measurement of the electric field.	
Different techniques have different strengths, such as speed, experimental simplicity and robustness. Another relatively recent approach is to use the pulse compressor itself, such as MIIPS~\cite{LozovoyOL2004}, chirp-scan~\cite{LoriotOE2013} and d-scan~\cite{MirandaOE2012}

Ultrashort pulse characterization techniques  that use the compressor itself as a diagnostic tool have the benefit of experimental simplicity, since minimal extra components are needed. These work by applying a set of distinct spectral phases, either using a pulse-shaping device or passive elements, and looking at the resulting second harmonic generation (SHG) spectra. These are usually scanning methods, but single-shot approaches have also been devised~\cite{FabrisOE2015}.

Such techniques retrieve the pulse iteratively, either by physically compressing the pulse~\cite{LozovoyOL2004} or by numerically finding which pulse originates the set of measured nonlinear (usually SHG) traces~\cite{LoriotOE2013,MirandaOE2012}. Our original implementation of the d-scan retrieval assumed an accurately measured fundamental spectrum, and tried to recreate the measured d-scan trace by iteratively applying spectral phases~\cite{MirandaOE2012,MirandaOE2012a}.	
Loriot et al. showed that the spectral intensity can also be retrieved from such traces~\cite{LoriotOE2013} if the trace is calibrated, in principle making a separate measurement of the fundamental spectrum unnecessary.	
In both cases, the pulse retrieval is treated as a generic optimization problem, where an iterative algorithm tries to find which pulse best recreates the measured scans. This is in contrast to some FROG algorithms~\cite{TrebinoJOSAA1993,DeLongOL1994,KaneJOSAB1997a,KaneIJSTQE1998}, which use more elegant approaches.

In this work, we present a new algorithm to retrieve ultrashort pulses from dispersion scans (d-scan). It is based on a generalized projections approach to phase recovery, similarly to what has been applied previously to FROG traces. The approach is inspired by decade-old techniques used in phase-recovery of diffraction problems, most famously the  Gerchberg--Saxton algorithm~\cite{GerchbergO1972}. In the context of ultrashort pulse measurement, this approach is used in most FROG retrieval algorithms~\cite{TrebinoJOSAA1993} as well as more recently developed algorithms, for example based on ptychography~\cite{SpangenbergOL2015,WittingOL2016}.

The algorithm iterates between a representation of the pulse (in the time or frequency domain) and its corresponding d-scan trace, i.e., its second-harmonic as a function of spectral phase. Calculating a d-scan trace from an ultrashort pulse is straightforward. However, and unlike linear problems such as diffraction or holography, finding the ultrashort pulse from a d-scan is not straightforward, even if the complex d-scan trace (i.e., intensity and phase) is known. The electric field generated from the SHG process is modeled as the square of the driving field; therefore, even if the SHG field is known, there is ambiguity to which driving field generated it. This step is arguably the crucial one in other retrieval techniques used in FROG.

\section{Fundamental Concepts}

We define the unknown pulse as the complex field
\begin{equation}
\tilde{E}(\omega)=|\tilde{E}(\omega)|\exp[i\psi(\omega)]
\end{equation}
and the glass spectral phase as a function of frequency \(\omega\) and glass thickness \(z\) is modeled as a complex function
\begin{equation}	
\Phi(\omega,z)=\exp(izk(\omega)).	
\end{equation}
Applying the spectral phase of different thicknesses of glass corresponds to translating wedges in the experimental measurement of a d-scan trace.
The complex d-scan, which is a set of complex SHG spectra, is given by
\begin{equation}
\label{eq:Somegaz}
S(\omega,z)=\invF{\F{\tilde{E}(\omega)\Phi(\omega,z)}^2}
\end{equation}
where \(\F{}\) is the Fourier transform operator, and the corresponding ideally measured d-scan trace is
\begin{equation}
\label{eq:exp_d-scan}
I(\omega,z)=\left|S(\omega,z) \right|^2.
\end{equation}

By integrating a trace over the insertion \(z\) from the minimum \(-L\) to the maximum glass insertion \(+L\)  we get the frequency marginal
\begin{equation}
M(\omega)=\int_{-L}^{+L} I(\omega,z) dz.
\end{equation}
We show in Appendix~\ref{app:marginals} that, for large enough values of \(L\), the marginal \(M(\omega)\)  is independent of the spectral phase of the pulse. It is thus a useful quantity that allows us to calibrate a d-scan trace.

Despite being a straightforward to implement technique, several experimental problems can arise. Depending on the particular case, it might be difficult to measure a SHG signal that follows the simple model of Eq.~\ref{eq:exp_d-scan}, due to phase-matching bandwidth, spectrometer calibration, contamination by the fundamental spectrum in the case of octave-spanning spectra, etc. It is thus a major advantage if the algorithm can deal with traces that are not ideal.

To investigate this we consider three different scenarios:
\begin{enumerate}
	\item The d-scan trace is well calibrated, without any clipping \label{well_calibrated}
	\item The d-scan trace is not calibrated, but it is not clipped, and the fundamental spectrum is properly measured \label{not_calibrated}
	\item The d-scan trace is possibly not calibrated, it is clipped, and the fundamental spectrum is properly measured \label{clipped}
\end{enumerate}

Case~\ref{well_calibrated} is the simplest one to implement numerically, and does not require an independent measurement of the fundamental spectrum. It is also the most closely related to FROG and ptychographic algorithms. The algorithm is described in Section~\ref{section:basic}.	
Case~\ref{not_calibrated} requires an independent measurement of the fundamental spectrum. There are then two options: one is to use the marginals to calibrate the d-scan trace, and then we have again case \ref{well_calibrated}. The other option is to change the retrieval algorithm to project the retrieved fundamental spectrum at each iteration (keeping the retrieved phase and applying the measured amplitude), while adapting the measured d-scan trace.

Case~\ref{clipped} is the most complex one, where data from some spectral regions of the d-scan trace is missing or unreliable. In this case, the areas of the trace which are not considered reliable are discarded and marked as such. The algorithm then replaces those by the simulated scan generated by the current guess. Successful retrieval of such traces depend on the d-scan trace having sufficient redundancy. Both cases \ref{not_calibrated} and \ref{clipped} are discussed in Section~\ref{section:clipped}.

\section{The Basic Algorithm}
\label{section:basic}

The basic algorithm is used in case \ref{well_calibrated}, when the trace is calibrated.

We start with a guess, for iteration \(i=0\)
\begin{equation}
\tilde{E}_{i=0}(\omega).
\end{equation}

From the d-scan trace we can estimate the width of the fundamental spectrum, and assume a Gaussian spectrum with that width as a first guess. A MIIPS analysis can be used for a first approximation to the spectral phase~\cite{LozovoyOL2004}.

From this we simulate a matrix of the field after propagation through a thickness \(z\). A phase matrix, corresponding to the spectral phase introduced by this propagation is
\begin{equation}
\Phi(\omega,z)=\exp(izk(\omega))
\end{equation}
so after propagation, the original field will have the extra spectral phase term as a function of glass thickness
\begin{equation}
\label{eq:1st}
\tilde{U}_i(\omega,z)=\tilde{E}_i(\omega) \Phi(\omega,z)
\end{equation}
giving a corresponding time domain matrix
\begin{equation}
U_i(t,z)=\F{\tilde{U}_i(\omega,z)},
\end{equation}
where the Fourier transform operator is applied over \(\omega\). Then the SHG signal is calculated
\begin{equation}
U^{SHG}_i(t,z)=U_i^2(t,z)
\end{equation}
and a complex d-scan trace is calculated:
\begin{equation}
\label{eq:complex_d-scan}
S_i(\omega,z)=\invF{U^{SHG}_i(t,z)}.
\end{equation}
Now, the phase is kept and the amplitude substituted by the measured trace
\begin{equation}
\label{eq:projection}
S'_i(\omega,z)=S^{meas}(\omega,z)\exp[\arg(S_i(\omega,z))]
\end{equation}
where \(S^{meas}(\omega,z)=\sqrt{I^{meas}(\omega,z)}\).
At each \(z\), we calculate a new guess for the SHG field
\begin{equation}
U'^{SHG}_i(t,z)=\F{S'_i(\omega,z)}.
\end{equation}
The most challenging part is to get the fundamental field that gives rise to that SHG field. By multiplying each field \(U'^{SHG}_i(t,z)\) by the complex conjugate of \(U_i(t,z)\) (which was our best guess so far),
\begin{equation}
\label{eq:mult_cc}
P_i(t,z)=U'^{SHG}_i(t,z)U_i^* (t,z)
\end{equation}
we hopefully get a better new guess for the phase, but the amplitude is off. To correct this we take the cube root of the amplitude of this term, 
and the next guess for the field at \(z\) is
\begin{equation}
\label{eq:thirdroot}
U'_i(t,z)=|P_i(t,z)|^{1/3}\exp\{\arg[P_i(t,z)]\}.
\end{equation}
Now we go again to the frequency domain by Fourier transforming
\begin{equation}
\tilde{U}'_i(w,z)=\invF{U'_i(t,z)}
\end{equation}
and the phase from the glass is removed
\begin{equation}
\tilde{U}''_i(w,z)=\tilde{U}'_i(w,z)\Phi^* (\omega,z).
\end{equation}
Finally, a new guess for the field is obtained by averaging over all the guesses corresponding to each glass insertion \(z\)
\begin{equation}
\label{eq:avg_freq}
\tilde{E}_{i+1}(w)=\frac{1}{\Delta z}\int_z \tilde{U}''_i(w,z) dz
\end{equation}
and the loop is repeated (from Eq.~\ref{eq:1st}) until a specified convergence criterion is met.

\subsection{Error Calculation}
At each iteration, the rms difference between measured and simulated traces can be calculated. The error is given by
\begin{equation}
G^2 = \frac{1}{N_j N_k} \sum_{j,k}\left( I^{meas}(\omega_j,z_k)-\mu I^{retr}(\omega_j,z_k)\right)^2 
\end{equation} 
where \(\mu\) is the constant that minimizes the error \(G\), which can be found by differentiating \(G\) in respect to  \(\mu\):
\begin{equation}
\label{eq:minimize_error}
\mu= \frac{\sum_{j,k} I^{meas}(\omega_j,z_k) I^{retr}(\omega_j,z_k) }{ \sum_{j,k} I^{retr}(\omega_j,z_k) }.
\end{equation}


\begin{figure}[htbp]
	\centerline{\includegraphics[]{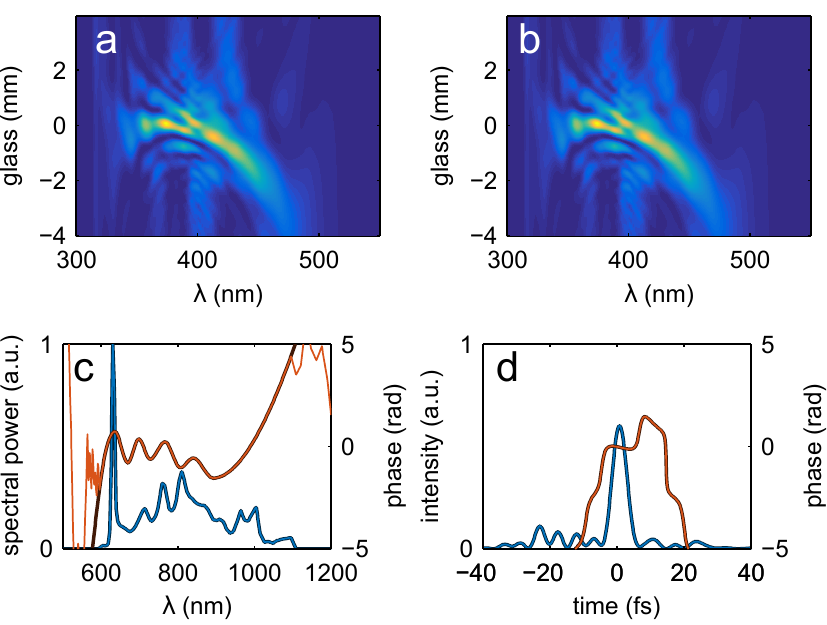}}
	\caption{Example of (a) simulated and (b) retrieved d-scan traces. (c) Both spectral intensity (blue) and phase (orange) are precisely retrieved, as well as the corresponding pulse in the time-domain (d). Retrieved fields are plotted in lighter colors. Note that in this plot the simulated and retrieved fields overlap almost exactly and cannot be easily distinguished.}
	\label{fig:no_noise}
\end{figure}	

\subsection{Simulated Examples}

Figure~\ref{fig:no_noise} shows an example of a simulated pulse and corresponding simulated and retrieved d-scan traces. The spectral intensity is an actual measurement of an ultrafast titanium:sapphire oscillator with a Fourier limit duration of less than 6~fs full width at half maximum (FWHM), representative of modern systems. This spectrum (amplitude and phase) will be used as a basis for more examples throughout this paper. The algorithm ran for 100 iterations, with a grid size of 256 points in the frequency domain and 200 glass insertions. 

The algorithm accurately retrieved the spectral intensity and phase. The RMS difference between the simulated and retrieved scans was of about 1E-6.

\begin{figure}[htbp]
	\centerline{\includegraphics[]{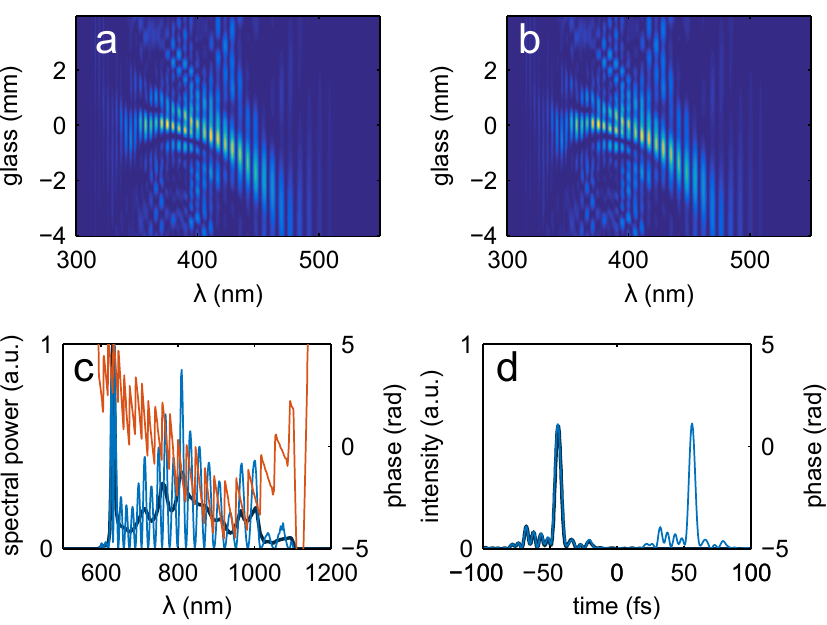}}
	\caption{Example of (a) simulated and (b) retrieved scans of a double pulse. The retrieved spectral intensity and phase are shown in (c) in light colors, as well as the corresponding spectral intensity for an isolated pulse. The corresponding pulses in the time domain are shown in (d).}
	\label{fig:double_pulse}
\end{figure}	

We also tested the basic algorithm on a double pulse (Fig.~\ref{fig:double_pulse}). The pulse is the same as the previous case (Fig.~\ref{fig:no_noise}), but duplicated and delayed by 100~fs. The only difference to the previous case was that a higher sampling rate was needed in the spectral domain (2048 points), which made the retrieval slower.

\subsection{Variations of the Basic Algorithm}

The previously presented basic algorithm can be used with some modifications. The first we consider is averaging the guesses in the time domain instead of doing it in the frequency domain. The step depicted in Eq.~\ref{eq:avg_freq} is replaced by its equivalent in time. Instead of integrating over \(\tilde{U}''_i(w,z)\), an extra step is taken: first, go to the time domain, integrate over \(z\), and go back to the frequency domain for the new \(\tilde{E}_i(w)\). The two approaches are equivalent, so if no other operation is performed there is no reason to do so. However, if some constraint is known in the time domain, it can be enforced here.

The other variation we tried is to give different weights to the different guesses. For example if the data is noisy, one might wish to give a higher weight to the regions with more signal. Equation \ref{eq:avg_freq} would then be replaced by
\begin{equation}
\tilde{E}_i(\omega)=\frac{\int_z \tilde{U}''_i(\omega,z) w(z) dz}{\int_z w(z) dz}
\end{equation}
where \(w(z)\) is the weight attributed to each spectral measurement, for example
\begin{equation}
w(z)=\int_{\omega}S^{meas}(\omega,z)d\omega.
\end{equation}
In the case of third-harmonic generation (THG) d-scan, the steps on Eqs.~\ref{eq:mult_cc} and \ref{eq:thirdroot} are replaced by:
\begin{equation}
P_i(t,z)= U'^{THG}_i(t,z)[U_i^* (t,z)]^2
\end{equation}
and
\begin{equation}
U'_i(t,z)=|P_i(t,z)|^{1/5}\exp\{\arg[P_i(t,z)]\}.
\end{equation}
These options will not be explored further in this work, and are mentioned only for completeness.

\section{Algorithm for Non-Calibrated Trace}
\label{section:clipped}

In the case of a non-calibrated trace, a retrieved trace will differ from the measured trace as~\cite{WeinerIJoQE1983,BaltuskaOL1998}
\begin{equation}
I^{meas}(\omega,z)=I^{retr}(\omega,z) R(\omega)
\end{equation}
where \(R(\omega)\) is a spectral response curve which includes factors as phase-matching bandwidth, spectrometer sensitivity, coupling efficiency, etc.
Since a simulated trace will be inherently calibrated, it will never recreate a non-calibrated measured trace, and the basic algorithm will not work properly.

One way around it is to use the trace's marginals~\cite{MirandaOE2012} to calibrate the trace. In some cases however (i.e. the trace was not measured with a large enough range of glass insertions) this might not be practical, so a similar approach to the one used in previous work~\cite{MirandaOE2012} can be adapted to the new algorithm.

A modified error expression is often used with d-scan retrieval~\cite{MirandaOE2012,MirandaOE2012a} which minimizes the error at each frequency
\begin{equation}
G^2 = \frac{1}{N_j N_k} \sum_{j,k}\left( I^{meas}(\omega_j,z_k)-\mu(\omega_j) I^{retr}(\omega_j,z_k)\right)^2.
\end{equation}
Similarly to Eq.~\ref{eq:minimize_error}, \(\mu(\omega_j)\) can easily be found by differentiation
\begin{equation}
\mu(\omega_j)= \frac{\sum_{k} I^{meas}(\omega_j,z_k) I^{retr}(\omega_j,z_k) }{ \sum_{k} I^{retr}(\omega_j,z_k) }.
\end{equation}
If the trace is not calibrated \emph{and} not clipped, then the retrieved response curve \(\mu(\omega_j)\) can be used to calibrate \(I^{retr}(\omega_j,z_k)\). It will only give the correct response after a successful retrieval, but if used within the retrieval algorithm, it will hopefully be a better guess at each iteration.

For a successful retrieval of a non-calibrated trace, a separate measurement of the fundamental spectrum is needed - this guarantees that the generated trace is itself calibrated.
Two extra steps are added to the algorithm. First, before a new d-scan guess is generated, the current guess of \(\tilde{E}_{i}(\omega)\) is updated with the measured spectrum, i.e., its phase is kept and its amplitude is replaced by the measured one:	
\begin{equation}
\label{eq:project_fundamental}
\tilde{E}'_{i}(\omega)= |\tilde{E}^{meas}(\omega)| \exp\{ \arg[ \tilde{E}_{i}(w) ] \} .
\end{equation} 
The d-scan guess is generated, and it is used to calibrate the measured trace. So an extra step is added, where the measured trace is replaced by the corresponding calibrated one
\begin{equation}
\label{eq:calibrate}
I^{meas}(\omega,z)'=I^{meas}(\omega,z)/\mu(\omega)
\end{equation}
where the current guess of \(I(\omega,z)\) (calculated from \(S(\omega,z)\) in Eq.~\ref{eq:complex_d-scan}) is used to calculate \(\mu(\omega)\).

In case some signal is missing or for some reason unreliable (for example contaminated by the fundamental spectrum in case of an octave-spanning pulse) those spectral regions cannot be used (case~\ref{clipped}). In this case, those areas are marked as unreliable and not substituted by the measured data in the projection step in Eq.~\ref{eq:projection}.

\begin{figure}[htbp]
	\centerline{\includegraphics[]{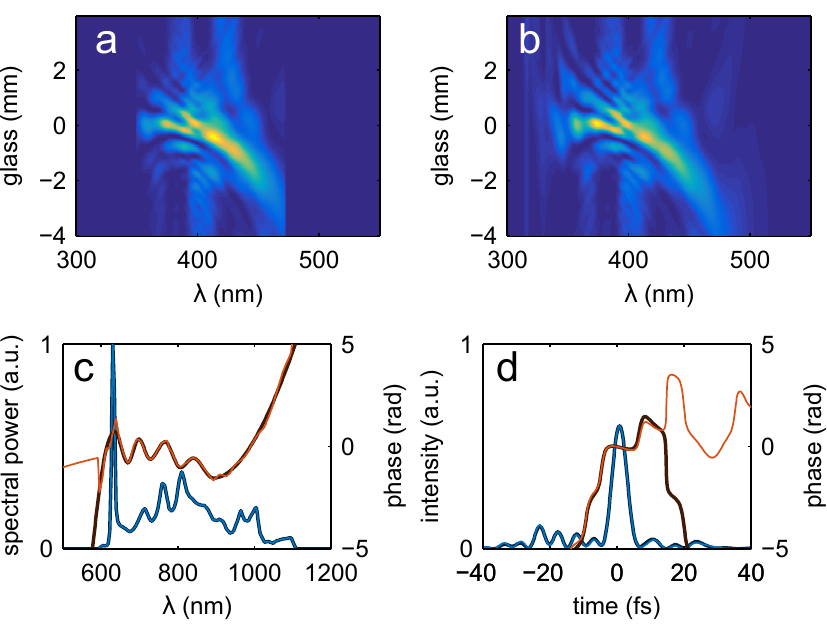}}
	\caption{Example of (a) simulated  trace where a spectral filter was applied, simulating phase-matching and clipping. The algorithm uses the known fundamental spectrum (c) and retrieves a trace (b) very similar to the original trace. The corresponding pulse in the time domain is shown in (d).}
	\label{fig:clipped_phase_mathing}
\end{figure}

Figure~\ref{fig:clipped_phase_mathing} shows an example using this variation of the algorithm, where the simulated trace was multiplied by a phase-matching curve, and spectrally clipped. Even in the absence of a significant portion of the trace, the pulse is accurately retrieved.

\section{Noise}
\begin{figure}[htbp]
	\centerline{\includegraphics[]{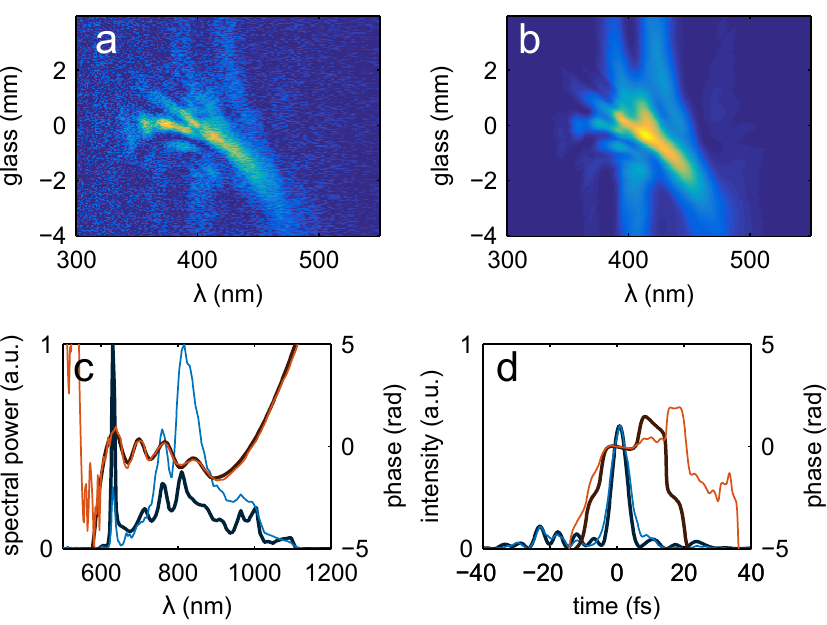}}
	\caption{Example of (a) simulated  trace with added noise and (b) corresponding retrieved trace. The basic algorithm retrieves (c) both spectral amplitude and phase, but especially the spectral amplitude (light-colored line) is not retrieved very accurately. In the time domain (d) the main features are reproduced but much less accurately than in the noiseless cases.}
	\label{fig:noise}
\end{figure}

We found that the basic algorithm is not particularly robust to noise. Figure~\ref{fig:noise} shows an example of a retrieval using the basic algorithm. Noise was added with a standard deviation of 10\% of the peak intensity of the trace. Especially the spectral intensity retrieval is strongly affected.

\begin{figure}[htbp]
	\centerline{\includegraphics[]{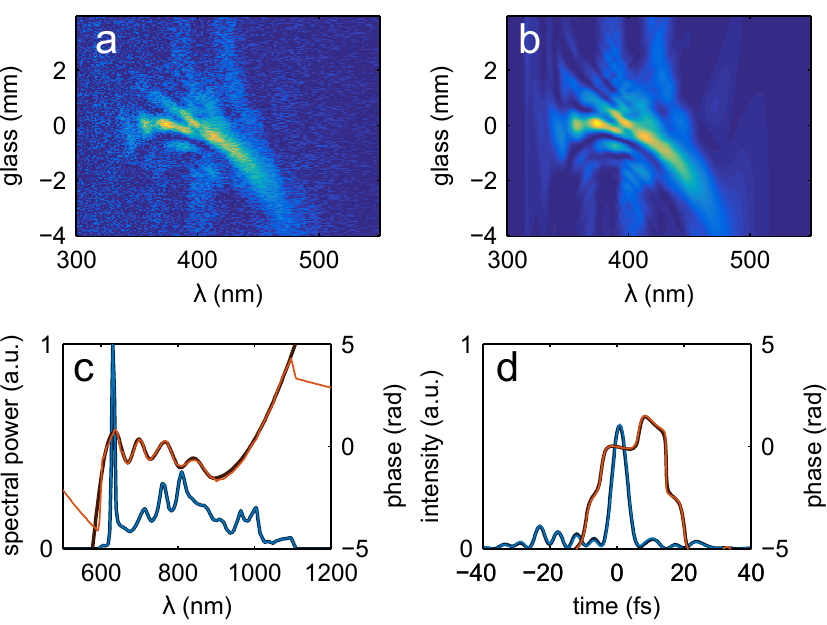}}
	\caption{Same as previous case, but with known fundamental spectrum. The algorithm variation that includes projection of the fundamental spectrum was used, yielding a much better reconstruction.}
	\label{fig:noise_proj_spect}
\end{figure}

If the fundamental spectrum is separately measured, then it can be integrated in the retrieval process as in the previous case (Eq.~\ref{eq:project_fundamental}). An example is shown in Fig.~\ref{fig:noise_proj_spect}.

\begin{figure}[htbp]
	\centerline{\includegraphics[]{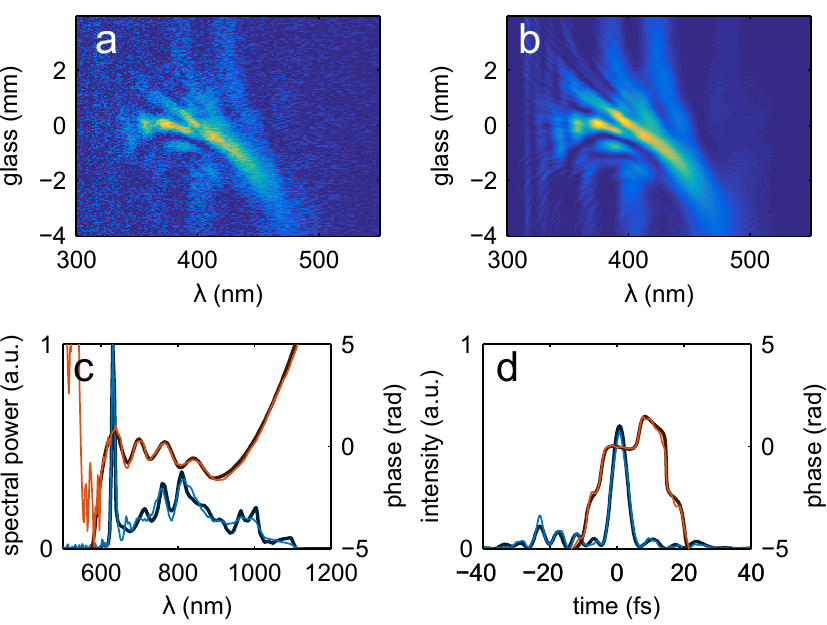}}
	\caption{Same as previous case, but using a mixed algorithm. After the retrieval using the basic algorithm, a multidimensional algorithm optimizes the spectral intensity. The retrieved spectral phase is left unchanged, and a generic optimization algorithm minimizes the error between (a) simulated and (b) retrieved scans by applying a filter curve to the previously retrieved spectral intensity.}
	\label{fig:noise_amp_opt}
\end{figure}

If the fundamental spectrum is not available, a composite algorithm can be used. From Fig.~\ref{fig:noise} we can see that the spectral phase retrieval is reasonably accurate, and this is very often the case even for very noisy traces. A simple approach is to take this retrieval and afterwards use a generic optimization algorithm to optimize the spectral intensity. Figure~\ref{fig:noise_amp_opt} shows an example.

A generic optimization algorithm takes the retrieved spectral amplitude from the basic algorithm and changes it, multiplying it by a response curve. In this case, the curve was defined by ten control points and interpolated using spline interpolation. This results in a much better retrieval, and can be done quite fast.

\subsection{Background Subtraction}

We finish this section by briefly discussing an important but often overlooked issue. When dealing with experimental data, the question arises on what to do with negative experimental values that arise from background subtraction.

In our previous work~\cite{MirandaOE2012,MirandaOE2012a}, we found that it is important \emph{not} to coerce to zero the experimental values of the d-scan trace, i.e, after subtracting the average noise value, negative values are kept negative, instead of being replaced by zero.

Let us assume we have an ideal trace, and look at a region where the values are zero. If noise with an average value of zero is added and then the resulting values are coerced to zero, the average will not be zero any longer. This would make the algorithm try to create signal in those regions (see for example Ref.~\cite{WilcoxJOSAB2014}). In the case of a generalized projections algorithm, where the measured \emph{amplitude} is needed, this poses a problem, as a square-root of the measured intensity is needed.

Our approach is to keep track of which values of the experimental trace are negative, and flip the sign of the complex scan when the projection step is performed. So the measured amplitude becomes
\begin{equation}
S^{meas}(\omega,z)=\sqrt{|I^{meas}(\omega,z)|}
\end{equation} 
and the projection step on Eq.~\ref{eq:projection} becomes
\begin{equation}
S'_i(\omega,z)=S^{meas}(\omega,z)\exp\{\arg[S_i(\omega,z)]\} \text{sgn}[I^{meas}(\omega,z)]
\end{equation}

We found this change to have a large impact on the retrieval of noisy traces.

\section{Experimental Results}

\subsection{OPCPA system}

\begin{figure}[htbp]
	\centerline{\includegraphics[]{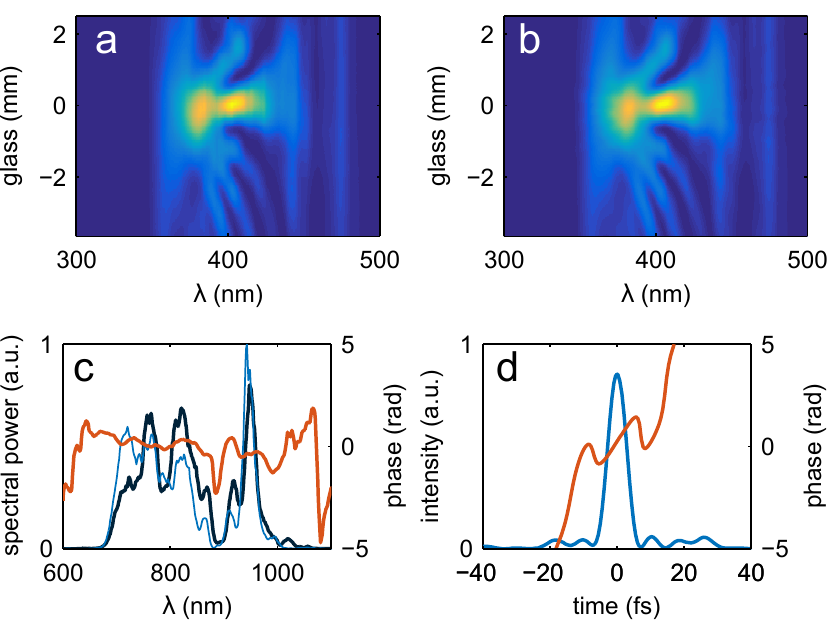}}
	\caption{Characterization of the output from an OPCPA system, with (b) retrieved trace very similar to (a) measured trace. (c) Retrieved spectral intensity and phase (light colored) are compared to the directly measured spectrum (dark color). (d) The shortest pulse (obtained at insertion 0) is 6.3~fs FWHM.}
	\label{fig:opcpa}
\end{figure}

The algorithm was tested on two different systems: the first is a chirped pulse optical parametric amplifier (OPCPA) co-developed with Venteon GmBH~\cite{RudawskiTEPJD2015}. It delivers around 5~\(\mu\)J of energy per pulse at a repetition rate of 200~kHz. The results are shown in Fig.~\ref{fig:opcpa}.

The measured pulses have a FWHM duration of 6.3~fs. In this case only the retrieval using the basic algorithm is shown. The retrieved spectrum is similar to the measured one, but the differences are clear. A possible explanation for the discrepancy is that either the fundamental or the SHG spectra are not properly calibrated. OPCPA sources are difficult to work with due to spatiotemporal couplings, making it hard to make sure that all the spectral content of the source contributes equally to the SHG signal. In any case, very similar results are obtained by instead using the previous algorithm~\cite{MirandaOE2012a} or the previously described variation for non-calibrated traces.

\subsection{Hollow-Fiber Compressor}

\begin{figure}[htbp]
	\centerline{\includegraphics[]{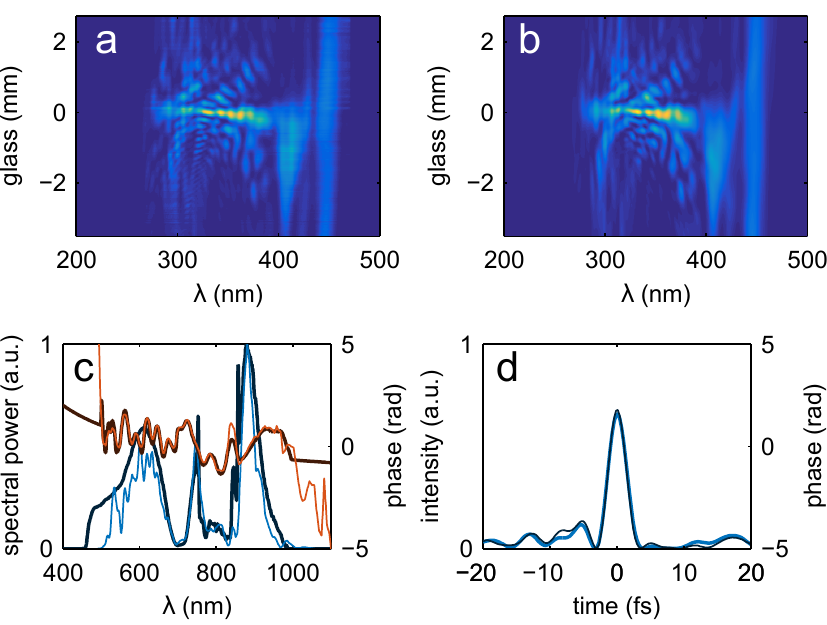}}
	\caption{Characterization of the output from a hollow-core fiber compressor. (a) Measured and (b) retrieved trace using the basic algorithm. Light colored lines correspond to the retrieval done with the basic algorithm (i.e. without using the measured spectrum). Dark colored lines correspond to the retrieval using the algorithm variation for non-calibrated traces, which uses the measured spectrum.  The retrieved pulse durations in both cases is 3.0~fs FWHM.}
	\label{fig:hollow_fiber}
\end{figure}

The second system is a hollow-fiber compressor~\cite{LouisyO2015} capable of delivering around 200~\(\mu\)J at a repetition rate of 1~kHz. In previous work, water was used to correct the third-order dispersion of the system~\cite{SilvaOE2014,LouisyO2015}. Here, we used a z-cut KDP crystal (so that light travels with polarization on the ordinary axis), which has similar TOD/GDD ratio to water. The results are shown in Fig.~\ref{fig:hollow_fiber}.
Two variations of the algorithm were used: the basic algorithm, and the modified algorithm using the measured spectrum and assuming the trace might not be calibrated. Both gave a very similar result, with a retrieved duration of 3.0~fs FWHM in both cases (the retrieved trace in Fig.~\ref{fig:hollow_fiber}b is the one obtained from the basic algorithm). The retrieved spectrum using the basic algorithm closely resembles the measured spectrum, except below 500~nm, which lies outside the chirped mirrors specs. This causes this spectral region to contribute very little to the main pulse, making it harder for the algorithm to retrieve it.

\section{Discussion}

The algorithm is not flawless, and sometimes it converges to local minima, and other times it doesn't converge at all. We found that a reasonable first guess helps in both cases. In the presented cases, it was enough to assume a Gaussian spectrum, transform-limited at the insertion corresponding to the peak of the d-scan signal as a first guess for the spectral phase.

The basic algorithm does not use all the information contained in a d-scan trace: the drop in intensity as a function of dispersion carries information which is not taken into account. Each spectral line is independently used for a new guess, so from the algorithm point of view the most important is that each spectral measurement is reproduced. If, for example, a trace had each spectral measurement normalized (which would give a physically impossible trace), the algorithm would still converge (although the error would not converge to zero).

Many of the steps of the algorithm can be done numerically in parallel, which can considerably speed up the retrieval. This is especially useful with modern multi-processor computers. 

\section{Conclusion}

We presented a new retrieval algorithm for ultrashort pulses using the d-scan technique, capable of retrieving both spectral intensity and spectral phase, and successfully applied it to simulated and real pulses. We also show that the frequency marginal of a d-scan trace is independent of the spectral phase of the pulse.

The algorithm can be implemented using different variations, depending on what data is available. It is much faster than our previous algorithms~\cite{MirandaOE2012,MirandaOE2012a} although, in its basic form, less robust to noise. 

The algorithm was tested on experimental setups, representative of some of today's most challenging ultrashort pulse measurements.

\textbf{Funding.} European Research Council (PALP), Knut and Alice Wallenberg Foundation, LASERLAB-EUROPE (grant agreement no. 654148, European Union’s Horizon 2020 research and innovation programme), Swedish Research Council, and Foundation for Science and Technology of Portugal (FCT).

\appendix

\section{Phase independence of the frequency marginal}
\label{app:marginals}
In this appendix we show that the integral of the d-scan trace $|S(\omega,z)|^2$ over the glass thickness $z$ is independent of the phase $\psi(\omega)$ of the pulse.
More precisely, we show that
\begin{equation}
\int_{-L}^{L}dz  |S(\omega,z)|^2 =  A(\omega)\log\frac{2L}{\ell(\omega)}+B(\omega)+\mathcal{O}(L^{-1})\,,
\label{eq:integralS2}
\end{equation}
where 
\begin{equation}
\ell(\omega)=\frac{ \tau^2}{
	\left|k''\left(\frac{\omega}{2}\right)\right|}
\,,\qquad
A(\omega)=\frac{1}{2\pi
	\left|k''\left(\frac{\omega}{2}\right)\right|} \left|\tilde{E}\left(\frac{\omega}{2}\right)\right|^4 \,,
\label{ellAomega}
\end{equation}
and
\begin{align}
&B(\omega)=\int \frac{d\omega_1}{\pi}
\label{eq:Bomega} \\ 
&\left[
\frac{\left|\tilde{E}\left(\frac{\omega}{2}+\omega_1\right)\right|^2
	\left|\tilde{E}\left(\frac{\omega}{2}-\omega_1\right)\right|^2}
{\left|k'\left(\frac{\omega}{2}+\omega_1\right)-k'\left(\frac{\omega}{2}-\omega_1\right)\right|}
- \frac{ \left|\tilde{E}\left(\frac{\omega}{2}\right)\right|^4 e^{-2\tau^2\omega_1^2}}{\left|2\omega_1 k''\left(\frac{\omega}{2}\right)\right|}
\right]\,.
\nonumber
\end{align}
In these equations, $\tau$ is an arbitrary time scale independent of $L$.
One can easily check that the derivative of Eq.~\ref{eq:integralS2} with respect to $\tau$ yields zero.

The derivation starts by rewriting Eq.~\ref{eq:Somegaz} for $S(\omega,z)$ as a convolution in the time domain
\begin{equation}
\int \frac{d\omega_1}{2\pi}
\tilde{E}\left(\frac{\omega}{2}+\omega_1\right)
\tilde{E}\left(\frac{\omega}{2}-\omega_1\right)e^{iz\left(
	k\left(\frac{\omega}{2}+\omega_1\right)+k\left(\frac{\omega}{2}-\omega_1\right)
	\right)}\,.
\end{equation}
Our goal is to compute the following triple integral in the limit of large $L$,
\begin{align}
&\int_{-L}^{L}dz  |S(\omega,z)|^2 =  \int \frac{d\omega_1 d\omega_2}{(2\pi)^2}
\int_{-L}^{L}dz
\,G(\omega,\omega_1,\omega_2,z)\,,
\label{tripleintegral}
\end{align}
where
\begin{align}
&G(\omega,\omega_1,\omega_2,z)
=e^{iz\left(
	k\left(\frac{\omega}{2}+\omega_1\right)+k\left(\frac{\omega}{2}-\omega_1\right)
	-
	k\left(\frac{\omega}{2}+\omega_2\right)-k\left(\frac{\omega}{2}-\omega_2\right)
	\right)}  \nonumber\\
&
\tilde{E}\left(\frac{\omega}{2}+\omega_1\right)
\tilde{E}\left(\frac{\omega}{2}-\omega_1\right)
\tilde{E}^*\left(\frac{\omega}{2}+\omega_2\right)
\tilde{E}^*\left(\frac{\omega}{2}-\omega_2\right)
\end{align}
If we naively set $L=\infty$ in Eq~\ref{tripleintegral}, the $z$-integral produces the following $\delta$-function,
\begin{equation}
\delta\left(
\tiny{k\left(\frac{\omega}{2}+\omega_1\right)+k\left(\frac{\omega}{2}-\omega_1\right)
	-
	k\left(\frac{\omega}{2}+\omega_2\right)-k\left(\frac{\omega}{2}-\omega_2\right)}
\right)\,.
\nonumber
\end{equation}
Assuming that $k(\omega)$ is a convex function, this $\delta$-function 
simplifies to
\begin{equation}
\frac{ \delta\left( \omega_2- \omega_1
	\right)+
	\delta\left( \omega_2+ \omega_1
	\right)}{\left|k'\left(\frac{\omega}{2}+\omega_1\right)-k'\left(\frac{\omega}{2}-\omega_1\right)\right|}\ .
\end{equation}
This simplifies Eq.~\ref{tripleintegral} to
\begin{align}
\int \frac{d\omega_1 }{ \pi }
\frac{\left|\tilde{E}\left(\frac{\omega}{2}+\omega_1\right)\right|^2
	\left|\tilde{E}\left(\frac{\omega}{2}-\omega_1\right)\right|^2}
{\left|k'\left(\frac{\omega}{2}+\omega_1\right)-k'\left(\frac{\omega}{2}-\omega_1\right)\right|}
\end{align}
which diverges from the region around $\omega_1=0$.
In order to tame this divergence, we write the integrand in Eq.~\ref{tripleintegral} as follows
\begin{align}
&G(\omega,\omega_1,\omega_2,z) 
- \left|\tilde{E}\left(\frac{\omega}{2} \right)\right|^4 e^{iz (\omega_1^2-\omega_2^2)k''\left(\frac{\omega}{2} \right)-\tau^2 (\omega_1^2+\omega_2^2)}\\
&+\left|\tilde{E}\left(\frac{\omega}{2} \right)\right|^4 e^{iz (\omega_1^2-\omega_2^2)k''\left(\frac{\omega}{2} \right)-\tau^2 (\omega_1^2+\omega_2^2)}\,.
\end{align}
The triple integral of the first line can now be computed directly in the limit $L\to \infty$ and it gives the finite result in Eq.~\ref{eq:Bomega}. On the other hand, the triple integral of the second line must be evaluated keeping $L$ finite. After performing the gaussian integrals over $\omega_1$ and $\omega_2$ we are left with the simple integral
\begin{align}
\frac{1}{2} A(\omega) \int_{-L}^{L} dz \frac{1}{\sqrt{z^2+\ell^2(\omega)}}=
A(\omega) \log\frac{L+\sqrt{L^2+\ell^2(\omega)}}{\ell(\omega)}\ , \nonumber
\end{align}
where we have used the definitions in Eq.~\ref{ellAomega}.
Finally, expanding at large $L$ we recover Eq.~\ref{eq:integralS2}.

\bibliography{Ref_lib}
\bibliographystyle{ieeetr}
\end{document}